\documentclass[12pt]{article}

\usepackage[english]{babel}
\usepackage[utf8]{inputenc}

\usepackage{doc}
\usepackage{amsmath}
\usepackage{amsfonts}
\usepackage{cmap}   
\usepackage{cite} 
\usepackage{fullpage} 
\usepackage[lmargin=0.6in,rmargin=0.6in,tmargin=0.6in,headsep=.2in]{geometry}
\usepackage{graphicx}
\usepackage{forloop}
\usepackage{etoolbox}
\usepackage{calc}
\usepackage{wrapfig,lipsum,booktabs} 
\usepackage{xtab} 
\usepackage[dvipsnames]{xcolor}
\usepackage[normalem]{ulem}
\usepackage{sectsty}
\usepackage{amsmath,amsfonts,amssymb,amsthm,epsfig,epstopdf,url,array}
\usepackage[retainorgcmds]{IEEEtrantools}
\usepackage{makeidx,epsfig,lscape}
\usepackage{xcolor,pict2e}

\usepackage[colorlinks=true]{hyperref}
\usepackage{doi}
\usepackage{orcidlink}

\usepackage{upgreek}
\usepackage{pgfplots}
\usepackage{float} 
\usepackage[sloped]{fourier}
\usepackage{comment}
\makeatother
\usepackage{makeidx}
\makeindex
\makeatletter
\renewcommand{\@biblabel}[1]{#1.}

\title{\bf Impact of Tariff Wars on Global Economy}

\author{\bf N.S. Gonchar \orcidlink{0000-0003-0954-8948}, O.P. Dovzhyk\orcidlink{0000-0003-4098-1192},A.S. Zhokhin\orcidlink{0000-0002-8969-3928},\\ \bf W.H. Kozyrsky\orcidlink{0000-0002-8969-3928}, A.P. Makhort\orcidlink{0000-0003-4098-1192}\\
\\ \href{https://bitp.kiev.ua}{\color{blue}Bogolyubov Institute for Theoretical Physics of NAS of Ukraine}}

\date{}

\begin{document}

\maketitle

\begin{abstract}

\noindent The Ricardian model of world trade based on comparative advantage is not sufficient to justify equal trade relations.The existing model of trade relations does not explain the distribution of income among trading countries. This paper presents a method for building equitable trade relations. Its essence is to present an algorithm for building such trade relations, based on the previously proposed model of world trade, that the trade balance of each country would be equal to zero.
Under such conditions, tariff wars would become impossible. It is proved that, provided that the supply structure is consistent with the demand structure, it is always possible to build an equilibrium price vector for which the trade balance of each country is zero. This state of economic equilibrium is called ideal. The article presents an algorithm to build an export structure based on the structure of imports. This algorithm is quite simple and allows for a wide range of applications.
Under fairly simple realistic assumptions about the behaviour of countries trading with each other that are subject to tariff restrictions, it is proved that this leads to an increase in the prices of the goods traded by these countries. Among the equilibrium states, there are also those called oversupply states. The latter describes the phenomenon of recession. This contributes to a fall in stock market indices.
\vskip 5mm

\end{abstract}

\section{Introduction.}

The model of international trade used by economists until now is based on Ricardo's ideas of comparative advantage, which only qualitatively explains the need for international trade. The Heckscher-Ohlin Theorem is a quantitative justification for trade between two countries by two types of goods~\cite{Heckscher1919,Ohlin1934,Lu2024}. Its further generalizations concerned increasing the number of goods traded between countries \cite{Samuelson1948,Krugman1991,Obstfeld2007,Yali2023,Rahman2022,Rubaj2022,Lewkowicz2024,Puslecki2024,Cui2023}.
Leontief's paradox showed the inability of this theory to explain the distribution of income in international trade. The purpose of this paper is to investigate the consequences of a tariff war between countries trading with each other. This analysis is based on the mathematical model of international trade proposed in the previous study \cite{Gonchar2022}. The model of international trade is based on the ideas and results obtained in monograph \cite{Gonchar2008}. In the paper \cite{Gonchar2022}, the existence of an equilibrium price vector at which there is a zero trade balance between trading countries is proved. An analysis of international trade between the most developed countries is presented in the  article \cite{	Gonchar2024}. Our model of international trade is based on the theory of economic equilibrium, that was further developed to describe economic systems under conditions of uncertainty. In the Theorem 1, under rather general assumptions about the structure of imports of trading countries, the value of exports of these countries is established at which there is a state of economic equilibrium such that the trade balance between countries is zero. This state is called the ideal state. In this state, markets are completely cleared. All other equilibrium states are described in a paper to be published in the journal: \textit{\textbf{Cybernetics and Systems Analysis, Vol. 61, N 3, 2025}}. The Theorem 2 shows that under fairly simple realistic assumptions about the behaviour of countries subject to tariff restrictions, this leads to an overall increase in prices for internationally traded goods. The described behaviour of countries in response to tariff restrictions will provoke a drop in indices on the stock exchanges of the respective countries. Note that in this paper we have considered only the state of economic equilibrium, which is ideal one. In other equilibrium states, tariff restrictions can lead to a recession \cite{Gonchar2015}.

\section{Model of International Trade.}

In the world economic system, let $l$ countries trade between themselves by $n$ types of goods. Suppose that the $i$-th country exports to the $s$-th country the vector of goods $e^{is}\mathrm{=}\mathrm{\{}e^{is}_k{\}}^n_{k\mathrm{=}1},$ and let $i$-th country imports from the $s$-th country the vector of goods $i^{is}\mathrm{=}\mathrm{\{}i^{is}_k{\}}^n_{k\mathrm{=}1}$. Then the vector of export of the $i$-th country will be \\ $b_i\mathrm{=}\mathrm{\{}b_k{\}}^n_{k\mathrm{=}1}\mathrm{=}\displaystyle\sum_{s\mathrm{\neq }i}{e^{is}}\mathrm{=}\mathrm{\{}\displaystyle\sum_{s\mathrm{\neq }i}{e^{is}_k}{\}}^n_{k\mathrm{=}1}$ and the vector of import of the $i$-th country will be \\ $$C_i\mathrm{=}\mathrm{\{}C_k{\}}^n_{k\mathrm{=}1}\mathrm{=}\displaystyle\sum_{s\mathrm{\neq }i}{i^{is}}\mathrm{=}\mathrm{\{}\displaystyle\sum_{s\mathrm{\neq }i}{i^{is}_k}{\}}^n_{k\mathrm{=}1}.$$
Since $e^{is}\mathrm{=}i^{si},s,i\mathrm{=}\overline{1,l},$ where we put $e^{ii}\mathrm{=}i^{ii}\mathrm{=}0,\ i\mathrm{=}\overline{1,l},$ we obtain

\begin{equation}\label{eq_lab1_}
	\begin{array}{r}
		\displaystyle\sum^l_{s\mathrm{=}1}{C_s}\mathrm{=}\displaystyle\sum^l_{s\mathrm{=}1}{\displaystyle\sum^l_{i\mathrm{=}1}{i^{si}}}
		\mathrm{=}\displaystyle\sum^l_{s\mathrm{=}1}{\displaystyle\sum^l_{i\mathrm{=}1}{e^{is}}}\mathrm{=}
		\displaystyle\sum^l_{i\mathrm{=}1}{\displaystyle\sum^l_{s\mathrm{=}1}{e^{is}}}
		\mathrm{=}\displaystyle\sum^l_{i\mathrm{=}1}{b_i}.
	\end{array}
\end{equation}

The vector of demand of the $i$-th country is determined by the vector $C_i\mathrm{=}\mathrm{\{}C_k{\}}^n_{k\mathrm{=}1}$ and supply vector is defined by the vector $b_i\mathrm{=}\mathrm{\{}b_k{\}}^n_{k\mathrm{=}1}.$ The equilibrium price vector $p_0\mathrm{=}\mathrm{\{}p^0_k{\}}^n_{k\mathrm{=}1}$ under which there is clearing of the market is determined from the set of equations

\begin{equation}\label{eq_lab2_}
\begin{array}{r}
\displaystyle\sum^l_{i\mathrm{=}1}{C_{ki}}
\frac{\mathrm{\langle}b_i,p_0\mathrm{\rangle }}{\mathrm{\langle }
C_i,p_0\mathrm{\rangle }}\mathrm{=}
\displaystyle\sum^l_{i\mathrm{=}1}{b_{ki}},\qquad k\mathrm{=}\overline{1,n},
\end{array}
\end{equation}

where~we~put $\mathrm{\langle }b_i,p_0\mathrm{\rangle }\mathrm{=}\displaystyle\sum^n_{s\mathrm{=}1}{b_{si}}p^0_s,\ \mathrm{\langle }C_i,p_0\mathrm{\rangle }\mathrm{=}\displaystyle\sum^n_{s\mathrm{=}1}{C_{si}}p^0_s,\ i\mathrm{=}\overline{1,l}.$

\textit{Tariff Restrictions}

Let $l$ countries have signed trade agreements with each other for a certain period of time. Suppose that during the period of validity of the agreements, some countries want to restrict access to their market for a significant number of goods by imposing tariff restrictions.

We assume that the vector of demand of the $i$-th country is determined by the vector $C_i\mathrm{=}\mathrm{\{}C_k{\}}^n_{k\mathrm{=}1}$ and supply vector is defined by the vector $b_i\mathrm{=}\mathrm{\{}b_k{\}}^n_{k\mathrm{=}1}$ and the equilibrium price vector $p_0\mathrm{=}\mathrm{\{}p^0_k{\}}^n_{k\mathrm{=}1}$ under which there is clearing of the market it is determined from the set of equation \eqref{eq_lab2_} and it is such that $\mathrm{\langle }b_i,p_0\mathrm{\rangle }\mathrm{-}\mathrm{\langle }C_i,p_0\mathrm{\rangle }\mathrm{=}0,i\mathrm{=}\overline{1,l}.$

We assume that three assumptions relative to trading between countries are valid:

\begin{enumerate}
\item  A reduction in revenue from trade between countries leads to a decrease in the production of goods for export.
\item  Symmetry in tariff restrictions between trading countries takes place.
\item  Symmetry in countries' actions regarding tariff restrictions leads to a reduction in both exports and imports, that depends solely on the type of goods.
\end{enumerate}

Let us clarify the assumptions made. As for the first assumption, it is obvious.
Indeed, the imposition of tariff restrictions by one country on another leads to a decrease in the revenue received from exports.

And this, in turn, leads to a decrease in the purchase of the amount of goods necessary for the production of goods for export.

The second assumption is also obvious. The imposition of tariff restrictions by one country entails a corresponding reaction from the other country.

The third assumption is essential and means that tariff restrictions lead to a reduction in exports or imports of goods that is independent of the countries trading with each other. Rather, it is a simplification in order to arrive at the desired effect of tariff restrictions in a simple way. In fact, this assumption means that the production technology of all countries trading with each other is the same.

Suppose that the $i$-th country imposed tariff restrictions on the $s$-th country. This means that if the $s$-th country was selling the $k$-th good at $p_k$, then after the tariff restriction it will sell it at $t^{is}_kp_k\mathrm{,\ }0\mathrm{<}t^{is}_k\mathrm{\le }1.$ If $t^{is}_k\mathrm{<}1,$ then we say that the tariff restriction on $k$-th product is imposed. The question arises how this changes the vector $i^{is}.$ Owing to first assumption, the vector $i^{is}$ should change to the vector $ri^{is}\mathrm{=}\mathrm{\{}r^{is}_ki^{is}_k{\}}^n_{k\mathrm{=}1},\ 0\mathrm{<}r^{is}_k\mathrm{\le }1,\ k\mathrm{=}\overline{1,n}.$ Let's take into account that $r_1e^{is}\mathrm{=}ri^{si}\mathrm{=}\mathrm{\{}r^{si}_ki^{si}_k{\}}^n_{k\mathrm{=}1}\mathrm{=}\mathrm{\{}r^{si}_ke^{is}_k{\}}^n_{k\mathrm{=}1}.$ Due to the third assumption $r^{si}_k\mathrm{=}r_k,\ k\mathrm{=}\overline{1,n}$ does not depend on $s,i,$ then tariff transformation of the vectors $ri^{is}\mathrm{=}\mathrm{\{}r_ki^{is}_k{\}}^n_{k\mathrm{=}1}\mathrm{,\ }re^{is}\mathrm{=}\mathrm{\{}r_ke^{is}_k{\}}^n_{k\mathrm{=}1}.$ From this we obtain the following transformations $rC_i\mathrm{=}\mathrm{\{}r_kC_{ki}{\}}^n_{k\mathrm{=}1},rb_i\mathrm{=}\mathrm{\{}r_kb_{ki}{\}}^n_{k\mathrm{=}1},\ i\mathrm{=}\overline{1,l}.$ Then the equilibrium price vector $p\mathrm{=}\mathrm{\{}p_k{\}}^n_{k\mathrm{=}1}$ should satisfy to the set of equations

\begin{equation}\label{eq_lab3_}
\begin{array}{r}
\displaystyle\sum^l_{i\mathrm{=1}}{r_k}C_{ki}\frac{\mathrm{\langle }rb_i,p\mathrm{\rangle }}{\mathrm{\langle }rC_i,p\mathrm{\rangle }}
\mathrm{=}\displaystyle\sum^l_{i\mathrm{=1}}{r_k}b_{ki}\mathrm{,}\mathrm{\qquad }k\mathrm{=}\overline{\mathrm{1,}n}.
\end{array}
\end{equation}

\section{Construction of an Equilibrium State\\ in the World Market}

Let us introduce the matrix $B\mathrm{=|}b_{ki}|^{n,l}_{k\mathrm{=}1,i\mathrm{=}1}$ composed from the vectors $b\mathrm{=}\mathrm{\{}b_{ki}{\}}^n_{k\mathrm{=}1}$ as columns and the matrix $C\mathrm{=|}C_{ki}|^{n,l}_{k\mathrm{=}1,i\mathrm{=}1}$ composed from the vectors $C_i\mathrm{=}\mathrm{\{}C_{ki}{\}}^n_{k\mathrm{=}1}$ as columns. Further, we assume that for the matrix $B\mathrm{=|}b_{ki}|^{n,l}_{k\mathrm{=}1,i\mathrm{=}1}$ the representation $B\mathrm{=}CB_1$ is true, where

\begin{equation} \label{eq_lab4_}
\begin{array}{c}
C\mathrm{=|}C_{ks}{\mathrm{|}}^{n,l}_{k,s\mathrm{=1}}\mathrm{,}\mathrm{\qquad }B_{\mathrm{1}}\mathrm{=|}b^{\mathrm{1}}_{ks}{\mathrm{|}}^l_{k,s\mathrm{=1}}, \\
b^{\mathrm{1}}_{ks}\mathrm{=}\displaystyle\sum^n_{i\mathrm{=1}}{\frac{{\tau }_{ki}C_{is}}{\displaystyle\sum^l_{s\mathrm{=1}}{C_{is}}}}\mathrm{,}\mathrm{\qquad }k,s\mathrm{=}\overline{\mathrm{1,}l}
\end{array}
\mathrm{\ }
\end{equation}

In the next Theorem 1 we prove the existence of the strictly positive solutions of a certain set of equations \eqref{eq_lab2_}.

Theorem 1. \textit{Let the matrix }${|t}_{ij}|^n_{i,j\mathrm{=}1}$\textit{= }$\mathrm{|}\displaystyle\sum^l_{k\mathrm{=}1}{C_{ik}}{\tau }_{kj}|^n_{i,j\mathrm{=}1}$\textit{ be a non negative and indecomposable one. Suppose that the matrix }$\tau \mathrm{=|}{\tau }_{kj}|^{l,n}_{k\mathrm{=}1,j\mathrm{=}1}$\textit{ is such that}

\begin{equation}\label{eq_lab5_}
\begin{array}{r}
\displaystyle\sum^n_{i\mathrm{=1}}{{\tau }_{ki}}\mathrm{=1,}\mathrm{\qquad }k\mathrm{=}\overline{\mathrm{1,}l}.
\end{array}
\end{equation}

\textit{Then the problem}

\begin{equation} \label{eq_lab6_}
\begin{array}{r}
\displaystyle\sum^l_{k\mathrm{=1}}{b^{\mathrm{1}}_{ks}}d_k\mathrm{=}d_s\mathrm{,}\mathrm{\qquad }s\mathrm{=}\overline{\mathrm{1,}l},
\end{array}
\end{equation}

\textit{has a strictly positive solution, belonging to the cone generated by columns of the matrix }$C^T,$\textit{ under conditions that }$\displaystyle\sum^l_{i\mathrm{=}1}{C_{ki}}\mathrm{>}0,\ k\mathrm{=}\overline{1,n},$\textit{ where }$C^T$\textit{ is transposed matrix to the matrix }$C.$\textit{ There exists an equilibrium price vector solving the set of equation (2).}

\textit{Proof.} Let us consider the nonlinear map

\begin{equation}\label{eq_lab7_}
\begin{array}{r}
H\left(p\right)\mathrm{=}\mathrm{\{}H_i\left(p\right){\}}^n_{i\mathrm{=}1}, \end{array}
\end{equation} 
\begin{equation}\label{eq_lab8_}
\begin{array}{r}
H_i\left(p\right)\mathrm{=}\frac{p_i\mathrm{+}\frac{\displaystyle\sum^n_{j\mathrm{=1}}{p_j}\displaystyle\sum^l_{k\mathrm{=1}}{C_{jk}}{\tau }_{ki}}{\displaystyle\sum^l_{s\mathrm{=1}}{C_{is}}}}{\mathrm{1+}\displaystyle\sum^n_{i\mathrm{=1}}{\frac{\displaystyle\sum^n_{j\mathrm{=1}}{p_j}\displaystyle\sum^l_{k\mathrm{=1}}{C_{jk}}{\tau }_{ki}}{\displaystyle\sum^l_{s\mathrm{=1}}{C_{is}}}}}\mathrm{,}\mathrm{\qquad }i\mathrm{=}\overline{\mathrm{1,}n},
\end{array}
\end{equation}
 
on the set $P\mathrm{=}\mathrm{\{}p\mathrm{=}\mathrm{\{}p_i{\}}^n_{i\mathrm{=}1},\ p_i\mathrm{\ge }0,i\mathrm{=}\overline{1,n},\ \displaystyle\sum^n_{i\mathrm{=}1}{p_i}\mathrm{=}1\}.$ It is continuous on $P$ and maps it into itself. Therefore, there exists a fixed point $p_0$ of the map $H\left(p\right)$ \cite{Nirenberg1974}. The last comes to the set of equations
\begin{equation} \label{eq_lab9_}
\begin{array}{r}
\frac{\displaystyle\sum^n_{j\mathrm{=1}}{p^0_j}\displaystyle\sum^l_{k\mathrm{=1}}{C_{jk}}{\tau }_{ki}}{\displaystyle\sum^l_{s\mathrm{=1}}{C_{is}}}\mathrm{=}\lambda p^0_i\mathrm{,}\mathrm{\qquad }i\mathrm{=}\overline{\mathrm{1,}n},
\end{array}
\end{equation}

where

\[\lambda \mathrm{=}\displaystyle\sum^n_{i\mathrm{=1}}{\frac{\displaystyle\sum^n_{j\mathrm{=1}}{p^0_j}\displaystyle\sum^l_{k\mathrm{=1}}{C_{jk}}{\tau }_{ki}}{\displaystyle\sum^l_{s\mathrm{=1}}{C_{is}}b^{\mathrm{1}}_s}}.\]

Let us prove that $\lambda \mathrm{>}0,$ and $p_0$ is a strictly positive vector. Let us introduce into consideration the matrix
\begin{equation} \label{eq_lab10_} 
\begin{array}{r}
F\mathrm{=|}f_{ji}|^n_{j,i\mathrm{=}1}, \end{array}
\end{equation} 
where
\[f_{ji}\mathrm{=}\frac{\displaystyle\sum^l_{k\mathrm{=1}}{C_{jk}}{\tau }_{ki}}{\displaystyle\sum^l_{s\mathrm{=1}}{C_{is}}}\mathrm{,}\mathrm{\qquad }i,j\mathrm{=}\overline{\mathrm{1,}n}.\]

The set of equations \eqref{eq_lab9_} can be written in the operator form
\[F^Tp_0\mathrm{=}\lambda p_0.\]

Since $F^T$ is non negative and indecomposable and the fact that the vector $p_0$ also solves the set of equations
\[{\left[F^T\right]}^{n\mathrm{-}1}p_0\mathrm{=}{\lambda }^{n\mathrm{-}1}p_0,\] 
we have that ${\left[F^T\right]}^{n\mathrm{-}1}p_0$ is a strictly positive vector.
From this it follows that $p_0$ is a strictly positive vector and $\lambda \mathrm{>}0.$ 
Here we introduced the denotation $F^T$ for the transposed matrix to the matrix $F.$ Prove that $\lambda \mathrm{=}1.$ Let us denote
\begin{equation} \label{eq_lab11_}
\begin{array}{r}
d_k\mathrm{=}\frac{1}{\lambda }\displaystyle\sum^n_{u\mathrm{=}1}{p^0_u}C_{uk},\qquad k\mathrm{=}\overline{1,l}.\end{array}
\end{equation}

Then, from \eqref{eq_lab9_} we obtain
\begin{equation} \label{eq_lab12_} 
\begin{array}{r}
p^0_i\mathrm{=}\frac{\displaystyle\sum^l_{k\mathrm{=}1}{d_k}{\tau }_{ki}}{\displaystyle\sum^l_{s\mathrm{=}1}{C_{is}}}.
\end{array}
\end{equation} 

Substituting \eqref{eq_lab12_} into \eqref{eq_lab11_}, we have
\begin{equation}\label{eq_lab13_}
\begin{array}{r}
d_j\mathrm{=}\frac{1}{\lambda }\displaystyle\sum^n_{u\mathrm{=}1}{\frac{\displaystyle\sum^l_{k\mathrm{=}1}{d_k}{\tau }_{ku}C_{uj}}{\displaystyle\sum^l_{s\mathrm{=}1}{C_{us}}}},\qquad j\mathrm{=}\overline{1,l}. 
\end{array}
\end{equation}

Or,
\begin{equation}\label{eq_lab14_}
\begin{array}{r}
d_j\mathrm{=}\frac{1}{\lambda }\displaystyle\sum^l_{k\mathrm{=}1}{\displaystyle\sum^n_{u\mathrm{=}1}{\frac{d_k{\tau }_{ku}C_{uj}}{\displaystyle\sum^l_{s\mathrm{=}1}{C_{us}}}}},\qquad j\mathrm{=}\overline{1,l}. \end{array}
\end{equation}

Summing up by index $j$ left and right parts of \eqref{eq_lab14_}, we obtain
\begin{equation}\label{eq_lab15_}
\begin{array}{r}
\displaystyle\sum^l_{j\mathrm{=}1}{d_j}\mathrm{=}\frac{1}{\lambda }\displaystyle\sum^l_{j\mathrm{=}1}{d_j}.
\end{array}
\end{equation} 

Since $\displaystyle\sum^l_{j\mathrm{=}1}{d^1_j}\mathrm{\neq }0,$ we have $\lambda \mathrm{=}1.$ Therefore, the system of equations \eqref{eq_lab13_} can be written in the form
\begin{equation}\label{eq_lab16_}
\begin{array}{r}
d_i\mathrm{=}\displaystyle\sum^n_{u\mathrm{=1}}{\frac{\displaystyle\sum^l_{k\mathrm{=1}}{d_k}{\tau }_{ku}C_{ui}}{\displaystyle\sum^l_{s\mathrm{=1}}{C_{us}}}}\mathrm{=}\displaystyle\sum^l_{k\mathrm{=1}}{d_k}\displaystyle\sum^n_{u\mathrm{=1}}{\frac{{\tau }_{ku}C_{ui}}{\displaystyle\sum^l_{s\mathrm{=1}}{C_{us}}}}\mathrm{,}\mathrm{\qquad }i\mathrm{=}\overline{\mathrm{1,}l}.
\end{array}
\end{equation}

Or,

\begin{equation}\label{eq_lab17_}
\begin{array}{r}
\displaystyle\sum^l_{k\mathrm{=}1}{b^1_{ki}}d_k\mathrm{=}d_i,\qquad i\mathrm{=}\overline{1,l}.
\end{array}
\end{equation}

Let us prove the solution existence of the set of equations \eqref{eq_lab2_}. Taking into account that
\begin{equation}\label{eq_lab18_}
\begin{array}{r}
d_k\mathrm{=}\displaystyle\sum^n_{u\mathrm{=}1}{p^0_u}C_{uk}\mathrm{=}\mathrm{\langle }C_k,p_0\mathrm{\rangle },\qquad k\mathrm{=}\overline{1,l}.
\end{array}
\end{equation} 

and the set of equations \eqref{eq_lab17_} we have

\begin{equation} \label{eq_lab19_} 
\begin{array}{r}
\mathrm{\langle }b_k,p_0\mathrm{\rangle }\mathrm{=}\mathrm{\langle }C_k,p_0\mathrm{\rangle },\qquad k\mathrm{=}\overline{1,l}. 
\end{array}
\end{equation}

From this it follows that the vector $p_0$ solves the set of equations \eqref{eq_lab2_}. Theorem 1 is proved. $\mathrm{\square}$

Theorem 2. \textit{Let }$l$\textit{ countries trade }$n$\textit{ types of goods, and let their trade are characterized by import and export vectors }$C_i,b_i,\ i\mathrm{=}\overline{1,l}$\textit{ where the import vector }$C_i\mathrm{=}\mathrm{\{}C_{ki}{\}}^n_{k\mathrm{=}1},i\mathrm{=}\overline{1,n},$\textit{ satisfies the conditions }$\displaystyle\sum^l_{k\mathrm{=}1}{C_{jk}}\mathrm{>}0,\ j\mathrm{=}\overline{1,n}.$\textit{ If the matrix }$\tau \mathrm{=|}{\tau }_{ki}{\mathrm{|}}^{l,n}_{k,i\mathrm{=}1}$\textit{ is such that the matrix }$t\mathrm{=|}t_{ji}{\mathrm{|}}^n_{i,j\mathrm{=}1}$\textit{ is nonnegative and indecomposable and then for the export vector }$b_i\mathrm{=}\mathrm{\{}b_{ki}{\}}^n_{k\mathrm{=}1}$\textit{ such that}
\begin{equation}\label{eq_lab20_}
\begin{array}{r}
b_{ik}\mathrm{=}\displaystyle\sum^n_{j\mathrm{=1}}{\frac{t_{ij}C_{jk}}{\displaystyle\sum^l_{r\mathrm{=1}}{C_{jr}}}}\mathrm{,}\mathrm{\qquad }i\mathrm{=}\overline{\mathrm{1,}n}\mathrm{,\ }k\mathrm{=}\overline{\mathrm{1,}l},
\end{array}
\end{equation}

\textit{the equilibrium price vector of the set of equations}

\begin{equation}\label{eq_lab21_}
\begin{array}{r}
\displaystyle\sum^l_{i\mathrm{=1}}{r_k}C_{ki}\frac{\mathrm{\langle }rb_i,p\mathrm{\rangle }}{\mathrm{\langle }rC_i,p\mathrm{\rangle }}\mathrm{=}\displaystyle\sum^l_{s\mathrm{=1}}{r_k}b_{ks}\mathrm{,}\mathrm{\qquad }k\mathrm{=}\overline{\mathrm{1,}n},
\end{array}
\end{equation}

\textit{exists and it is equal to the vector }$p\mathrm{=}\mathrm{\{}\frac{p^0_i}{r_i}{\}}^n_{i\mathrm{=}1},$\textit{ where the vector }$p_0\mathrm{=}\mathrm{\{}p^0_i{\}}^n_{i\mathrm{=}1}$\textit{ solves the set of equations}

\begin{equation}\label{eq_lab22_}
\begin{array}{r}
\displaystyle\sum^l_{i\mathrm{=1}}{C_{ki}}\frac{\mathrm{\langle }b_i,p_0\mathrm{\rangle }}{\mathrm{\langle }C_i,p_0\mathrm{\rangle }}\mathrm{=}\displaystyle\sum^l_{s\mathrm{=1}}{b_{ks}}\mathrm{,}\mathrm{\qquad }k\mathrm{=}\overline{\mathrm{1,}n}.
\end{array}
\end{equation}

Consequence 1. \textit{Tariff wars between trading countries leads to the increasing of prices in international trade if }$r_i\mathrm{<}1,\ i\mathrm{=}\overline{1,n}.$

\section{ Conclusion.}

The paper proposes an algorithm to build fair international trade. It is based on the model of international trade proposed in the article \cite{Gonchar2022}. To do this, we find the ratio between the exports and imports of trading countries, for which there is such an ideal equilibrium that the trade balance between all countries is zero. Under simple realistic assumptions about the behaviour of countries subject to tariff restrictions, it is proved that this leads to an increase in prices for goods traded between countries. In a less than ideal equilibrium, tariff restrictions can lead to a recession.

\section{Conflicts of Interest.}

The authors declare no conflicts of interest regarding the publication of this paper.

\noindent \textbf{}

\section{Funding.}

This work is partially supported by the Fundamental Research Program of the Department of Physics and Astronomy of the National Academy of Sciences of Ukraine "Building and researching financial market models using the methods of nonlinear statistical physics and the physics of nonlinear phenomena N 0123U100362".


\vskip 5mm

\noindent 
https://orcid.org/0000-0003-0954-8948 (Nicholas Simon Gonchar),\\ mhonchar@i.ua
\orcidlink{0000-0003-0954-8948}

\noindent
https://orcid.org/0000-0003-4098-1192 (Olena Petrivna Dovzhyk),\\ alenkadov87@gmail.com
\orcidlink{0000-0003-4098-1192}

\noindent
https://orcid.org/0000-0001-7826-6608 (Anatoly Sergiyovych Zhokhin),\\ aszhokhin@gmail.com
\orcidlink{0000-0001-7826-6608}

\noindent
https://orcid.org/0000-0002-8969-3928 (Wolodymyr Hlib Kozyrski),\\ kozyrski@ukr.net
\orcidlink{0000-0002-8969-3928}

\noindent
https://orcid.org/0000-0003-4098-1192 (Andrii Pylypovych Makhort),\\ map@bitp.kiev.ua
\orcidlink{0000-0003-4098-1192}

\end{document}